\documentclass[aps,jcap,nofootinbib,groupedaddress,amsfonts,floatfix]{revtex4}

\usepackage{graphicx,amsmath,amssymb,amstext}
\usepackage{amssymb,amsbsy,amsfonts,amsthm,hyperref}
\usepackage{lscape,placeins}
\usepackage{url}
\usepackage[normalem]{ulem}

\newcommand{\adsurl}[1]{\href{#1}{ADS}}
\providecommand{\url}[1]{\href{#1}{#1}}

\newcommand{\be}{\begin{equation}}
\newcommand{\ee}{\end{equation}}
\newcommand{\bea}{\begin{eqnarray}}
\newcommand{\eea}{\end{eqnarray}}

\usepackage{color}
\usepackage{ifthen}
\newboolean{editorial}
\setboolean{editorial}{true}
\newcommand{\editorial}[2]{\ifthenelse{\boolean{editorial}}{\textcolor{red}{[\textsf{\textbf{{#1}}}: }\textcolor{blue}{\textsf{{#2}}}\textcolor{red}{]}}{}}

\definecolor{amber}{rgb}{1.0, 0.49, 0.0}

\begin{document}

\title{What is flat $\Lambda$CDM, and may we choose it?}

\author{Stefano Anselmi ${}^{1,2}$}
% \email[]{stefano.anselmi@pd.infn.it}
\author{Matthew F. Carney${}^{3}$}
\author{John T. Giblin, Jr${}^{4,5}$}
\author{Saurabh Kumar${}^{4}$}
\author{James B. Mertens${}^{3}$}
\author{Marcio O'Dwyer${}^{4}$}
\author{Glenn D.~Starkman${}^{4,6}$}
\email[]{glenn.starkman@case.edu}
\author{Chi Tian${}^{3,7}$}

\affiliation{${}^1$INFN-Padova, Via Marzolo 8, I-35131 Padova -- Italy}
\affiliation{${}^2$ LUTH, UMR 8102 CNRS, Observatoire de Paris, PSL Research University, Universit\'e Paris Diderot, 92190 Meudon -- France}
\affiliation{${}^3$Department of Physics and McDonnell Center for the Space Sciences, Washington University, St. Louis, MO 63130, USA}
\affiliation{${}^4$Department of Physics/CERCA/Institute for the Science of Origins, Case Western Reserve University, Cleveland, OH 44106-7079 -- USA}
\affiliation{${}^5$Department of Physics, Kenyon College, 201 N College Rd, Gambier, OH 43022, USA}
\affiliation{${}^6$Department of Physics, Imperial College, London, UK}
\affiliation{${}^7$School of Physics and Optoelectronics Engineering, Anhui University, Hefei,China}

\date{\today}

\begin{abstract}
The Universe is neither homogeneous nor isotropic, but it is close enough that we can reasonably approximate it as such on suitably large scales.
The inflationary-$\Lambda$-Cold Dark Matter ($\Lambda$CDM) concordance cosmology builds on these assumptions to describe the origin and evolution of fluctuations. With standard assumptions about stress-energy sources, this system is specified by just seven phenomenological parameters,
whose precise relations to underlying fundamental theories are complicated and may depend on details of those fields. 
Nevertheless, it is common practice to set the parameter that characterizes the spatial curvature, $\Omega_K$, exactly to zero.
This parameter-fixed $\Lambda$CDM is awarded distinguished status as separate model, ``flat $\Lambda$CDM.''  {\it Ipso facto} this places the onus on proponents of ``curved $\Lambda$CDM'' to present sufficient  evidence that $\Omega_K\neq0$, and is needed as a parameter.
While certain inflationary model Lagrangians, with certain values of their parameters, and certain initial conditions, will lead to a present-day universe well-described as containing zero curvature, this does not justify distinguishing that subset of Lagrangians, parameters and initial conditions into a separate model.
Absent any theoretical arguments, we cannot use observations that suggest small $\Omega_K$ to enforce $\Omega_K=0$.
Our track record in picking inflationary models and their parameters {\it a priori} makes such a choice dubious, and
concerns about tensions in cosmological parameters and large-angle cosmic-microwave-background anomalies strengthens arguments against this choice.
We argue that $\Omega_K$ must not be set to zero, and that $\Lambda$CDM remains a phenomenological model with at least 7 parameters.
\end{abstract}

\maketitle

Concordance cosmology, a Universe that is dominated by dark energy of fixed energy density and cold dark matter ($\Lambda$CDM) and that began, long ago, with a nearly scale-free spectrum of adiabatic density fluctuations, is
a surprisingly simple description of a remarkably complex universe.
In this setup, just about everything we see on cosmological scales
can be described in terms of 
a nearly homogeneous and isotropic space-time 
containing
homogeneous and isotropic vacuum energy, and filled with
ordinary baryonic and dark matter, photons, and neutrinos
that are nearly homogeneous on cosmological scales, 
but have sizable fluctuations on astrophysical scales.
The dynamics of this energy density is governed by general relativity (GR) and the Standard Model of particle physics (and possibly some other interactions if the dark matter is not of the Standard Model), though the evolution of cosmological perturbations can be tracked almost entirely with linearized GR. 
In addition, one commonly postulates 
an early-universe ``inflationary'' origin for the  fluctuations, which explains why they are primordially adiabatic,
with a nearly scale-invariant power spectrum over some wide range of scales.

This cosmology can be parametrized in terms of just 7 real parameters:
$\Omega_b h^2$,
$\Omega_c h^2$,
$H_0$,
$\tau$,
$n_s$,
$A_s$,
and $\Omega_K$.\footnote{
One might add $T_{\rm CMB}$, the current temperature of the cosmic microwave background radiation (CMB), which fixes the cosmological moment at which we are describing the universe. 
$T_{\rm CMB}$ also fixes the current energy density in radiation $\Omega_r$ once one has specified $N_{\rm eff}$, the effective number of relativistic neutrino species at the time of big bang nucleosynthesis. 
$N_{\rm eff}$ is calculable in the Standard Model.
These parameters are those of what is often called ``vanilla'' $\Lambda$CDM.
There are also a number of additional parameters that characterize detailed features of the Universe or departures from the simplest model. 
These include the neutrino masses and mixings, and the amplitude and spectral parameters of any tensor modes.
There are many other such potential  parameters.}
In more detail, these are\footnote{Throughout, we set $c=1$.}:  
	the ratio $\Omega_b$ of the energy density in baryons $\rho_b$ to the critical energy density, $\rho_{crit} = 3H_0^2/8\pi G_N$, for a homogeneous flat Friedmann-Lema\^{i}tre-Robertson-Walker (FLRW) universe, times $h^2$, the square of the current Hubble parameter $H_0$ in units of $100$ km/s/Mpc;
	$h^2$ times $\Omega_c$, 
    the ratio of the energy density in cold dark matter $\rho_c$ to $\rho_{crit}$;
    the current Hubble parameter $H_0$;
    the reionization optical depth $\tau$ (which should in principal be a calculable function of the other 6, but in practice depends on complicated non-linear astrophysics);
    the index $n_s$ of the power spectrum of scalar fluctuations, and the amplitude $A_s$ of that power spectrum.
    Finally, there is $\Omega_K$, a measure of spatial curvature, written to appear like the ratio of the energy density in spatial curvature to the $\rho_{crit}$.  Actually there is no energy density associated with spatial curvature, which is part of the geometric (Einstein tensor $G_{\mu\nu}$) term of the Einstein field equations $G_{\mu\nu}+\Lambda g_{\mu\nu}=8\pi G_N T_{\mu\nu}$, not the source (stress-energy tensor $T_{\mu\nu}$) term.  However, in an FLRW universe, $\Omega_K$ is $1$ minus the sum of the $\Omega_i$ associated with each of the actual sources of energy density -- once the cosmological constant $\Lambda$ has been rewritten as a vacuum-energy-density contribution to $T_{\mu\nu}$.

It is noteworthy that all  these  parameters are phenomenological, i.e. not one can be readily understood as a parameter of the action of the fundamental theory underlying cosmology.  
They are more closely analogous to the masses, orbital parameters and similar properties of the Sun and planets in a ``theory'' of the solar system,  than to General Relativity's only parameters, $G_N$ and the value of the cosmological constant $\Lambda$, or to the 19-20 parameters of the Standard Model of particle physics (plus a few more if one includes neutrino masses and mixings). 
The  $\Lambda$CDM parameters depend in complicated, and in some cases not-yet-understood, ways not just on the parameters of the fundamental theory--which is not yet fully known--but also, at least potentially, on boundary conditions. 
A rough analogy is  to the dynamics of a plasma inside a cavity, which depend not just on the micro-physics of the plasma constituents, but also on the shape of the cavity, the conductivity of its surface, and even the roughness of that surface.
It should therefore perhaps be surprising that just 7 parameters seem to suffice to approximately describe the Universe.

It has nevertheless become the widespread practice among cosmological phenomenologists and observers to regard the last of these, $\Omega_K$, as optional; to set $\Omega_K\equiv0$ and analyze the ``6-parameter flat-$\Lambda$CDM'' concordance cosmology. This argument rests on three legs:
First, the are the 
historical low-precision measurements of $\Omega_K$, from which  there  was  no evidence  that  $\Omega_K\neq0$.
Second, there are long-standing arguments that $\Omega_K=0$ is a generic prediction of inflation, though, during a period in the 1990s when $\Omega_m\simeq0.3$ was the preferred observational value, and it was believed that $\Omega_\Lambda\equiv0$, so $1-\Omega_m-\Omega_\Lambda=0.7=\Omega_K$, some  argued that $\Omega_K\neq0$ was generic, or at least \cite{Linde:1995xm} that there was no good reason to prefer any particular value of $\Omega_K$.   These arguments have largely not been revisited  since  the  emergence  of  tensions in the fits of $\Lambda$CDM parameters to cosmological data, and of statistical anomalies in that data.
Finally, third, there is a  sense among cosmologists that there are three choices -- $k=0$, $k=1$, and $k=-1$ -- and that these are separate models that could be treated separately, even though this is a choice not between distinct fundamental Lagrangians but between boundary conditions (i.e. between three classes of possible manifolds, that don't even exhaust the full set of possible manifolds).

We set out here to challenge this practice of regarding $(\Omega_K=0)$-$\Lambda$CDM as a distinct model with one fewer parameter than $\Lambda$CDM, which can be evaluated separately.
It has been suggested that it is ``premature'' \cite{Leonard:2016evk} to set $\Omega_K=0$ in $\Lambda$CDM, and argued that this is a ``very restrictive assumption
that needs to be tested empirically'' \cite{Planck:2015fie}, 
leading many others to explore $\Lambda$CDM with $\Omega_K\neq0$ (e.g., \cite{Planck:2018vyg,DiValentino:2019qzk,Handley:2019tkm,Efstathiou:2020wem}) with a range of conclusions.
Still others have worried about the possibility of drawing erroneous conclusions about evolving dark energy \cite{Clarkson:2007bc,Wright:2006kx},  the nature of inflation \cite{Handley:2019anl}, or alternatives to inflationary $\Lambda$CDM \cite{Melia:2018kyb}.
Here we argue that one simply is not entitled set $\Omega_K=0$ in $\Lambda$CDM.
By downgrading an inescapable phenomenological parameter of the model universe to an optional parameter,
we introduce--through our priors--information that we simply don't have {\it a priori}.
By insisting that the parameter should be restored only if the need for it is justified by strong evidence, we potentially mislead ourselves about tensions among data sets, how they might be solved, our confidence in our concordance model, the utility of future and planned experiments, and other key questions in the era of precision cosmology. The point of this paper is not to take sides for or against particular tensions, but to argue that the correct stage on which they should be examined cannot be assumed to be flat -- 
Without a theory of initial fluctuations, such as inflation, indentifying
$\Omega_K$ indistinguishable from $0$ as a ``model'' relies completely on arbitrary  assumptions about the properties of cosmic fluctuations to even make sense of a background cosmology, much less select the specific Euclidean background.
Within the inflationary $\Lambda$CDM paradigm, it is a prediction of certain, but not all, inflationary Lagrangians, with certain, but definitely not all, values of those Lagrangian's parameters, and certain, but again not all, initial conditions -- all in a context where there is no well-understood measure on the models/parameter/initial-condition ``space" from which one is choosing.

Properly including $\Omega_K$ as a free parameter  has practical consequences: changing the best-fit values of other parameters and their uncertainties, and  exacerbating or moderating the strengths of parameter tensions. 
In principle, one may also wish to  replace $\Omega_K$ with one or more underlying model parameters that can be more closely connected to calculable predictions for observables.

\section{What does it mean to be flat?}

The subtext of this question is that FLRW cosmology equates/conflates at least three distinct meanings of
the statement ``the Universe is flat'':
\begin{enumerate}
    \item The Universe is a Euclidean three-manifold:  All of spacetime may be foliated by spacelike hypersurfaces each of which is a Riemannian three-manifold (an appropriately smooth space) with the appropriate topology to ``admit'' a single homogeneous metric, and that is the Euclidean metric. This property might extend back to just after a spacelike ``Big Bang'' singularity, or to $t=-\infty$ in a cyclic or bouncing cosmology; but most importantly, it includes the time slice  we inhabit.
    \item The observable Universe has zero average curvature: the ``three-geometry'' of the constant-time spacelike hypersurface through us is characterized by a 3-Riemann-curvature tensor $^3R_{ijkl}$. The spatial average of this 3-geometric quantity (or the 3-Ricci tensor, or the 3-Ricci scalar) is $0$ over the region of space that we observe.
    \item The (first) Friedmann equation requires no term that scales like curvature: cosmological observations are analyzed in the context of a phenomenological model---background FLRW cosmology plus perturbations. The background has only one dynamical degree-of-freedom, the scale factor $a(t)$.  The  Friedmann equation, supplemented with an equation-of-state relationship between pressure and energy density, relates the rate of change of $a(t)$, to the energy density $\rho$.
    In ``vanilla'' $\Lambda$CDM, four contributions to $\rho$ are included, which scale as\footnote{
        When placed in the context of the Standard Model of particle physics, or various extensions thereto, specific particles contribute to either the matter or radiation component at different epochs.  As the Universe expands,  particles with mass transition from radiation-like to matter-like.  This slightly complicates the simple four-component description, and is of particular current relevance for neutrinos.} $a^{-4}$  (``radiation''), 
        $a^{-3}$ (``matter''), $a^{-2}$ (``curvature''\footnote{
        In General Relativity, the presumed underlying theory of FLRW, an $a^{-2}$ term can have a geometric origin, arising from an isotropic homogeneous curvature of space. It thus enters the Einstein equations through the Einstein tensor, not the stress-energy tensor. Certain forms of energy density, such as certain networks of  cosmic strings, would also scale as $a^{-2}$, but these are assumed not to be present in vanilla $\Lambda$CDM.
    }), 
    and $a^0$ (``vacuum energy'').  
    The current fractional contributions to the total energy density are given by $\Omega_i$, with $i$ indexing these four sources. (Thus $\sum_i\Omega_i\equiv 1$.) 
    $\Omega_K$ is the symbol for the fractional coefficient of the $a^{-2}$ contribution to the Friedmann equation.  
\end{enumerate}

The word flat is thus non-specific.   
Our not-precisely homogeneous or isotropic universe may or may not be flat in each of the above three senses. 

We examine each of these meanings of flatness below, the goal being to ask if, and to what extent, we are justified in assuming that any or all of the above flatness conditions are satisfied {\it a priori}, i.e. what are the implications of placing a strict prior that perfectly enforces any or all of these conditions on the  cosmology, and which then requires strong evidence to overturn.

The question is not, for example, how good or bad an approximation it is to expand around a homogeneous, isotropic, zero-spatial-curvature background, but rather whether one can justify regarding perfectly flat FLRW as a separate model, in which, in particular, $\Omega_K$ is simply not a parameter that one must measure.

\subsection{Topological  flatness}
An underlying assumption of FLRW is that $3+1$-dimensional classical (i.e. ignoring any underlying quantum nature)
spacetime is smooth\footnote{
    We try to steer clear of specific mathematical definitions when those are not necessary.}---a semi-Riemannian manifold---and can be described as a time-ordered sequence of smooth $3$-d spatial slices, at least from soon after some possible initial singularity to the present and into the foreseeable future.
We know that this is a problematic assumption given the existence of black holes, which are thought to hide singularities, or quantum-mechanically resolved singularities, behind horizons. 
Nevertheless, we have the sense, and assume, that we can excise these short-distance defects in the geometry. 

We also assume for concordance cosmology that this large-scale spatial 3-manifold has no boundary hypersurfaces.
This is not an essential feature of spacetime;
in certain models of large extra dimensions in which interest exploded in the 1990's
\cite{Arkani-Hamed:1998jmv,Randall:1999ee,Kaloper:2000jb},
we (i.e. all or most Standard Model particles) are confined to  ``the brane''--a 3+1-dimensional sub-spacetime
of the full ``bulk'' spacetime.  
This brane generically imposes boundary conditions on bulk fields.
Nevertheless, a no-boundary prior is defensible, if not essential.

Having assumed that space on large scales is a three-manifold,  we confront their vast diversity.
Spatial three-manifolds can be infinite (non-compact) or finite (compact).
Compact manifolds without boundary are called closed\footnote{
    This meaning of a closed manifold should not be confused with the cosmologists' description of a universe that eventually stops expanding and then collapses toward a future big crunch as a ``closed universe.''
   $S^3$ LFRW universes are closed in both senses if $\Lambda$ is sufficiently small that curvature causes the cosmic expansion to halt before $\Lambda$ can cause accelerated expansion.
    } 
and have been extensively studied.
Many cosmologists are familiar with the fact that in two dimensions
every simply connected Riemann surface admits one of just three homogeneous geometries -- Euclidean ($E^2$), spherical ($S^2$), or hyperbolic ($H^2$).
Effectively, it is possible to smooth out any local ripples in the spacetime and leave behind a homogeneous and isotropic geometry (i.e. metric).

The situation is much less simple in three dimensions.
Thurston, in 1982, put forward the eponymous Geometrization Conjecture  \cite{Thurston1982ThreeDM}, since proved by Perelman \cite{perelman2002entropy,perelman2003ricci}. 
It states, approximately, that every closed 3-manifold is the connected sum of irreducible pieces called prime 3-manifolds. 
This decomposition is unique for orientable manifolds\footnote{
    We may or may not inhabit an orientable manifold.  The Standard Model is chiral -- it includes different left-handed and right-handed fermionic particles/fields.  On a non-orientable manifold this is not possible. On the other hand, it seems risky to infer global properties of spacetime from no-go theorems whose subtle assumptions may prove consequential.
    }.
Each prime submanifold admits a homogeneous geometry--however there are 8 of them, not 3.  
Non-compact three-manifolds in general are much less well studied.

FLRW considers only the three homogeneous and isotropic 3-geometries---Euclidean ($E^3$), spherical ($S^3$), and hyperbolic ($H^3$).
Because the dynamical equations of General Relativity depend only on the metric, the Einstein field equations for the evolution of the homogeneous metric (but not their boundary conditions) are entirely agnostic about which of the many manifolds of each type we might inhabit.
Cosmologists therefore tend to equate these geometries with the covering spaces of these geometries -- for each geometry, the manifold that has the full unbroken symmetry group of that geometry -- because the field equations of GR are independent of the topology.
Cosmologists thus refer to all $E^3$ manifolds as $k=0$, all $S^3$ manifolds as $k=1$, and all $H^3$ manifolds as $k=-1$.
Two more of the 8 possible homogeneous geometries follow quite naturally from the two-dimensional case: $S^2\times E^1$, and $H^2\times E^1$.
They are anisotropic---flat in one dimension and isotropically curved in the other two. 
The final three, $\widetilde{SL}(2, R)$, Nil, and Sol, are less familiar, and also anisotropic.

It is a challenge to place any measure on this rich space of possibilities. 
There are only $10$ topologically distinct closed 3-manifolds with the $E^3$ geometry.
On the other hand there are a countable infinity of distinct topologies for manifolds admitting the $S^3$ geometry.
Meanwhile ``there are enormous numbers of examples of ... [$H^3$ manifolds], and their classification is not completely understood.'' \cite{wikipediaGeometrizationconjecture}.
This suggests that flat 3-manifolds are a set of measure zero in the space of closed 3-manifolds.   
However, we should be cautious to adopt the topologists' counting for cosmological purposes.
For example, hyperbolic three-manifolds are rigid \cite{Mostow:1968} -- the only deformation of the fundamental domain permitted if one wants to preserve the homogeneity of the metric is an overall scaling (i.e. a change in the curvature scale).  
In contrast, the simplest Euclidean three-torus is usually described as a cube with opposite faces identified; however,
each of the three side lengths can be independently and continuously varied, as can the three independent angles between the sides. 
Thus the finite number of flat three-manifolds each require up to 6 real parameters to specify them, whereas the countable infinity of hyperbolic three-manifolds require only one each.  
On the other hand, it is not clear why we should count the parameters that specify the homogeneous geometry, 
and disregard the overwhelmingly larger set of real parameters required to specify inhomogeneities   -- represented, for example, by the amplitudes of eigenmodes of the appropriate Laplacian. 

Returning to and refining our initial assertion -- it is a challenge to select a single compelling measure for cosmology on the space of possible three-manifolds.\footnote{We avoid even contemplating the complications if the topology of the $3+1$d spacetime precludes a global decomposition into a spatial manifold cross time.}  

In summary, 
 the cosmologists' classification of FLRW models as $k=0,\,\pm1$ cannot be taken as a rational basis for assigning priors to three-manifolds, as it is grossly incomplete, and arbitrarily implies something like equal weight for each of  the three special classes of manifolds that it happens to include, and zero weight for all other three-manifolds, not to mention spaces that aren't manifolds, but that would/might be just as viable cosmologically.

 It is distinctly possible that the entire question of cosmic topology is obviated by a prolonged epoch of cosmic inflation. 
Suppose the pre-inflationary universe was homogeneous but with a geometry other than $E^3$. 
Or perhaps was nearly homogeneous, but well described by a background geometry other than $E^3$ plus perturbations.
During cosmic inflation, the length scale(s) associated with that (background) geometry would increase compared with the Hubble Horizon scale and any other scale determined by microphysics.   
Local measurements of geometry--i.e. measurements made over fixed distances--might 
be increasingly well-approximated by Euclidean geometry.\footnote{
    Cosmologists would normally say that local measurements of geometry \emph{would} be increasingly well-approximated by Euclidean geometry; however, that statement hides the important assumption of a short-distance cut-off in the spectrum of initial inhomogeneities -- perhaps the physical Planck scale at the onset of inflation.
}
After inflation ended, the physical scale over which such ``local'' measurements could be made would begin to increase--this is often described as scales coming (back) ``into the horizon.''
If inflation was sufficiently brief, then we could today, once again, detect the non-Euclidean nature of the background geometry.
If inflation lasted long enough, then even today, the deviation of the background geometry from Euclidean might be too small to detect.

Detecting the background curvature and detecting the non-triviality of cosmic topology are not one and the same thing,
though they are certainly connected.  
Moderately stringent limits have been placed on observables related to the topology of the Universe. 
The shortest closed loops ``around the Universe'' passing through us must, at 99.7\% CL, be longer than at least 
98.5\% of the diameter of the last scattering surface of the cosmic microwave background (CMB) \cite{Cornish:2003db,ShapiroKey:2006hm,Vaudrevange:2012da} -- a statement that can be made
regardless of the specific manifold in which we might live, and of the specific background geometry given the inferred levels of curvature fluctuaions. 
This is known based on the absence of ``matched circle pairs'' \cite{Cornish:1997ab} in Wilkinson Microwave Anisotropy Probe (WMAP) temperature maps \cite{WMAP_ILC_onLAMBDA}, slightly extended to special cases using Planck data \cite{Planck:2013okc,Planck:2015gmu}.
The Planck team also searched for evidence of non-trivial topology using a correlation-function technique, but that search was limited to certain classes of manifolds, to certain values of the parameters of the applicable fundamental domain, and, for most topologies, to certain locations within that fundamental domain.  
The possibility remains  of non-trivial topology on length scales short enough that it could be detected.

In summary, a spatial slice through us may have the appropriate topology to allow us to put the homogeneous Euclidean 3-metric on its entirety (once we somehow excise the black holes and ``stitch up'' those excisions).  This could reasonably be called a topologically flat space. 
It is hard (or impossible) to put a meaningful measure on the space of 3-spaces to evaluate the probability for this to be the case. 
We  have not  measured the topology (and may never do so), so do not yet have any basis for settling the question observationally. 
Neither do we have any basis from cosmic topology for assuming the Universe is flat.

\subsection{Zero average spatial curvature}

Whether or not the spatial slice through us (or at least a judiciously chosen slice) is a topologically flat 3-manifold (i.e. admits a homogeneous Euclidean metric), the actual metric is not homogeneous -- everywhere around us is evidence of non-Euclidean geometry.  Perhaps, however, the Universe is flat because the average spatial curvature over the observable Universe is zero.
There are at least three questions pertaining to this suggestion:
\begin{enumerate}
\item Which average of which quantity is zero?
\item Which theoretical models predict that that average will be precisely zero?
\item Is there justification for assigning a prior to one or more of those models that plucks them out of the vast space of models and distinguishes them among their alternatives?
\end{enumerate}

\subsubsection{Which average curvature?}
While the geometry of spacetime is described by a metric $g_{\mu\nu}$, 
the curvature is characterized by the Riemann tensor $R_{\alpha\beta\gamma\delta}$, which, written this way with 4 covariant indices, is a function of the second derivatives of the metric with respect to the spacetime coordinates. 
$R_{\alpha\beta\gamma\delta}$ has 
20 independent degrees of freedom. 
The Einstein field equations, 
the dynamical equations of GR, depend only on the Ricci tensor, a trace of the Riemann tensor  $R_{\mu\nu}=g^{\alpha\gamma}R_{\alpha\mu\gamma\nu}$ 
(and its trace, the Ricci scalar, $R=g^{\mu\nu}R_{\mu\nu}$), 
through the Einstein tensor\footnote{
    The remaining degrees of freedom of the Riemann tensor are contained in the Weyl or conformal tensor.
}
$G_{\mu\nu}=R_{\mu\nu}-\frac{1}{2} R g_{\mu\nu}$.  
$G_{\mu\nu}$ has just 10 independent components.
The field equations state that $G_{\mu\nu}$ is proportional to the stress-energy tensor $T_{\mu\nu}$,
\begin{equation}
    \label{eqn:EinsteinFieldEqns}
   G_{\mu\nu} = 8\pi G_N T_{\mu\nu}\,.
\end{equation}
For a universe that is close to homogeneous and isotropic, and which seems to be reasonably well-modeled by some mixture of 
perfect fluids (characterized by their density $\rho$ and isotropic pressure $p$), the diagonal elements of $T_\mu^\nu$ in a Cartesian coordinate system -- $\rho,p,p,p$ -- are most definitely not zero on average at most cosmological epochs -- during the radiation-dominated and matter dominated periods they are positive definite; during the current accelerated-expansion epoch (and any past inflationary epoch), $\rho$ is positive definite while $p$ is negative definite.
Thus none of $G_{\mu\nu}$, $R_{\mu\nu}$, and $R$ average to 0 over a cosmologically interesting volume, at least for any reasonable choice of average.

This is not surprising to any cosmologist.  ``Flat'' cannot mean that spacetime has no curvature -- it is a claim about space not spacetime,
in other words about the induced metric on spatial slices. 
We must first perform a ``space-time decomposition'' -- we must take the $3+1$-dimensional spacetime manifold and represent it as a sequence of non-intersecting $3$-dimensional spatial manifolds, a foliation.
On any given spatial slice any two points are outside one anothers' light-cones.  
The coordinates within any given slice are spatial coordinates, the label for the different slices is time.
On each such slice, we can then find\footnote{$^3g$ can be computed directly from $g$ and the specification of the foliation.}
 the metric $^3g_{ij}$ describing distances between pairs of points (with $i,j$ now running over 3 spatial coordinates),
and we can compute the 3-d spatial Riemann tensor, $^3R_{ijkl}$ from spatial derivatives of $^3g$. 
When cosmologists speak of a flat universe, they are describing the properties of $^3R_{ijkl}$.  
$^3R_{ijkl}$ has 6 degrees of freedom, but 
in the Euclidean 3-geometry $E^3$, $^3R_{ijkl}=0$.

We can immedately identify several challenges.
First, the separation of spacetime into space and time is not unique.
Even if there is a choice of foliations that gives $^3R_{ijkl}=0$, this will not be true on other foliations.
We could adopt a very mathematical approach and merely require that there exist such a foliation,
but how would one verify this observationally?  
And what precisely would one mean?  After all, the geometry is very far from Euclidean on small scales.  More on that below.

The second and related challenge is that we do not observe spatial slices.
The only point on our spatial slice that we know we can observe is the one we occupy. 
Mostly we observe photons (or lately gravitational waves and neutrinos) that emanate from points on (or incredibly near, in the case of neutrinos) our past light cone --
approximately these are  nested spheres of increasing radius as one moves further into the past.
(We do get information from our past worldline -- e.g. geological data; we also get photons that have scattered in the past, 
offering in principle an opportunity to fill in the light cone, but that is far less extensive.
Cosmic rays do this to a limited extent as well, travelling at all speeds $\leq c$.)
But this simple description relies on the Universe being nearly homogeneous. 
 The presence of mass concentrations, which create gravitational lenses, black holes, {\it{et cetera}}, complicates this description.
On which surface are we meant to average the curvature and find zero?  

Finally, what quantity are we meant to average and how?
Is it $^3R_{ijkl}$ or $^3R^i_{jk}l$; the 3-Ricci tensor $^3R_{ij}$ or $^3R^i_j$; or the 3-Ricci scalar, $^3R$?
Or is it some function of these?
The average of a function is not the function of the average; averaging doesn't commute with most mathematical operations.
And how do we weight the average? 

\indent{\it And is it zero?}
One clear challenge in  placing a delta-function prior on any average of the curvature at zero is that no viable theory predicts that the curvature will be precisely zero. 
This is absolutely clear on ``small scales'' -- the existence of structures from Cavendish experiments to superclusters testifies to the curvature of space. 
Most of these structures, while non-linear from the point of view of their stress-energy, source gravitational fields that can be studied at low order in perturbation theory; however this is certainly not the case for the mergers of stellar mass black hole binaries, which we now detect regularly via their gravitational radiation.
Thus any attempt to assert that the curvature is zero must involve a scheme to remove the contribution of these linear and non-linear deviations from an idealized flat background. 
Many authors
\cite{Zeldovich1964, Bertotti1966,Gunn1967a,Gunn1967b, Kantowski1969, DyerRoeder1972, DyerRoeder1974,Kaiser:2015iia, Bonvin:2015kea} 
have explored analytically the question of which averages of which observables might be best suited for recovering an unbiased measure of the background geometry, although the expected variance has not as far as we can tell been studied.
Recently,
Adamek {\textit et al.} \cite{Adamek:2018rru} explored the influence of small-scale non-linear structures on determinations of certain curvature averages and provided numerical evidence supporting the claim \cite{Kaiser:2015iia} that well-designed measurements could at least be unbiased. (See below.)

Meanwhile, current theories of the origins of structure ascribe the origin of the small scale structures we inhabit and observe to a spectrum of fluctuations in some primordial field, typically a scalar field termed the inflaton, with associated fluctuations in the metric.
This spectrum naturally extends to wavelengths much larger than the size of currently known structures, indeed larger than the Hubble scale.
Kleban \cite{Kleban:2012ph} argued that if the fluctuations in the Newtonian potential are $\Phi_N$,
then they generate  $^3R = \frac{4\nabla^2\Phi_N}{a^2}$.
(Of course there can be vector and tensor perturbations, anisotropic stress, and other non-Newtonian contributions.)

There is no compelling reason to expect a sudden cutoff in the spectrum at scales just larger than observed structures. 
For example, some of us \cite{Copi:2018wsv} demonstrated that such a cutoff, somewhat unexpectedly, does not naturally explain the absence of large angle correlations in the CMB temperature fluctuations. 
Cosmic topology would provide a cutoff, but only on scales larger than the fundamental domain, which, as described above, is already known not to be much smaller than the sphere of last scattering -- the limits of our direct electromagnetic observations.
Both the ``subhorizon'' and ``superhorizon'' contributions to the metric fluctuations would contribute to any average of the curvature on large scales. 
Kleban showed that for a canonical inflationary power spectrum, the square root of the expectation value of the spatial average of $(^3R)^2$ 
is non-zero.  
Interpreted as a background curvature it would lead to $\vert\Omega_K\vert \simeq 10^{-5}$.
The magnitude of the contribution of large-scale modes to the average curvature (as characterized by 
$\vert\Omega_K\vert$) was explored numerically in \cite{Adamek:2018rru} and by some of us in \cite{Tian:2020qnm}, where we also considered different averaging procedures.  
While long-wavelength fluctuations make a non-vanishing contribution to inferred cosmological parameters,  
for that contribution to be observationally significant
would require that their amplitude is larger than anticipated in  vanilla inflationary $\Lambda$CDM.
Exactly this has been suggested as a possible solution to the $H_0$ tension \cite{Adhikari:2019fvb}.    

This spatial average of $(^3R)^2$ may, anyway, not be an observationally relevant quantity to explore. 
Certainly, while $(^3R)^2=0$ implies $\Omega_K=0$, the converse is not true; it is not necessary to have $(^3R)^2=0$ in order to get $\Omega_K=0$.
We are interested in observables, such as the  coefficient of the $a^{-2}$ term in the Friedmann equation,
not an  integral over a gauge-dependent spatial hypersurface.
Moreover, a similar computation to those in \cite{Adamek:2018rru} or
\cite{Tian:2020qnm}
would presumably yield a non-isotropic effective curvature, probably of similar size to the average curvature, but which might be more detectable.

\subsection{$\mathbf{\Omega_{-2}=0}$ for scale-factor evolution}

In a homogeneous and isotropic universe, the metric in spherical polar coordinates can be written in the form
\begin{equation}
    \label{eqn:FRWmetric}
    ds^2 = dt^2 - a(t)^2\left[\frac{dr^2}{1-kr^2} + r^2\left(d\theta^2+\sin^2\theta d\phi^2\right)\right]
\end{equation}
where $k$ is the curvature parameter described above.
This metric is consistent with a homogeneous diagonal stress energy tensor of the form $T^\mu_\nu(t)=\mathrm{diag}(\rho(t),p(t),p(t),p(t))$.
The Einstein equations for the dynamics of the metric then reduce to a pair of independent equations for the 
first and second time derivatives of $a(t)$ in terms of the density $\rho(t)$, the pressure $p(t)$, 
and two constants, the cosmological constant $\Lambda$ and the curvature constant $k$. 
These are the Friedmann equations:
\begin{flalign}
    \label{eqn:Friedmann1}
    \left(\frac{\dot{a}}{a}\right)^2 &= \frac{8\pi G\rho}{3} - \frac{1}{3}\Lambda  - \frac{k }{a^2}\\
    \label{eqn:Friedmann2}
    \frac{\Ddot{a}}{a} &= -\frac{4\pi G\rho}{3}\left( \rho + 3 p \right) - \frac{1}{3}\Lambda\,.
\end{flalign}
Applied to a homogeneous isotropic perfect fluid with a given equation of state, $p=p(\rho)$, or to a collection of such fluids (i.e. $\rho=\sum_i\rho_i$),
the Friedmann equations yield the time evolution of the one dynamical function in \eqref{eqn:FRWmetric}--the scale factor $a(t)$.

We know that the contributions to the energy density in the Universe include certain elements--relativistic particles such as photons, massive slow-moving composite ``particles'' such as atoms, ions, and nuclei (and the still-more-composite objects built out of them like stars and galaxies).  
The latter category is widely understood to also include cold dark matter, non-relativistic non-interacting ``particles''.
A gas of photons, which are massless and non-self-interacting, is very well approximated as a perfect fluid with $p=\rho c^2/3$.
Because $p$ is the kinetic energy density of a perfect fluid, for non-relativistic non-interacting species, $p \simeq \frac{1}{2}\rho v^2 \ll \rho c^2$,  so $p=0$ is  a good leading order approximation.
This is known as ``dust.''

Conservation of stress-energy (or equivalently the Friedmann equations) imply that
a fluid with $p=\rho/3$ has $\rho\propto a^{-4}$, while one with $p=0$ has $\rho\propto a^{-3}$.
This allows the first Friedmann equation in a homogeneous isotropic universe containing only radiation ($p_r=\rho_r c^2/3$) and
pressureless matter to be rewritten as a first-order non-linear ordinary differential equation:
\begin{equation}
    \label{eqn:IntroduceOmegai}
    \left(\frac{\dot{a}}{a}\right)^2 =  H_0^2 \sum_{i=0,-2,-3,-4} \Omega_i \left( \frac{a}{a_0}\right)^{i}
\end{equation}
with
\begin{flalign}
    \label{eqn:Omegasubintegers}
    \Omega_0 &\equiv -\frac{\Lambda c^2}{3 H_0^2}\,, \quad
    \Omega_{-2} \equiv -\frac{k c^2}{a_0^2 H_0^2}\,, \quad
    \Omega_{-3} \equiv \frac{8\pi G}{3H_0^2}\rho_{m0}\,,\\ \nonumber
    &{\mathrm{and}} \quad
    \Omega_{-4} \equiv \frac{8\pi G}{3H_0^2}\rho_{r0}\,.
\end{flalign}
$H_0$ is $\dot{a}/a$ evaluated today, and $\rho_{m0}$ and $\rho_{r0}$ are the dust and radiation energy densities when $a=a_0$.
Conventionally, cosmologists replace $\Omega_0\to\Omega_\Lambda$, $\Omega_{-2}\to\Omega_K$,  $\Omega_{-3}\to\Omega_m$, $\Omega_{-4}\to\Omega_r$.

The framework described above relies implicitly on the confidence that the metric that solves the full Einstein equations sourced by the real stress energy tensor is well approximated by the homogeneous metric that is sourced by the average stress-energy tensor 
\begin{flalign}
    &\bar{g}_{\alpha\beta}\left[G_{\mu\nu}\!=\!8\pi G_N {T}_{\mu\nu}\right]\\ \nonumber 
    &\qquad\qquad 
    \simeq g_{\alpha\beta}\left[G_{\mu\nu}\!=\!8\pi G_N \bar{T}_{\mu\nu}\right] \,,
\end{flalign}
(where spatial averages over FLRW hypersurfaces are represented by an overbar);
and, moreover, that the average stress-energy tensor is well described by the sum of homogeneous perfect fluids that have the same equations of state as the constituent fluids would if they were actually homogeneous
\begin{flalign}
     &\bar{T}_{\mu\nu}
     \simeq T_{\mu,\nu}[\bar{\rho},p(\bar{\rho})]\,.
\end{flalign}

This casual commutation of averaging with the solving of systems of coupled non-linear partial differential equations is not to be taken lightly, and has long been the subject of concern under the name of ``the cosmlogical backreaction problem'' 
and remains unsettled (see, for example, \cite{Schander:2021pgt} and references therein for a recent review).
It raises several questions:
\begin{itemize}
    \item Is the average stress energy tensor well described as the sum of a homogeneous radiation fluid and a homogeneous dust fluid?  Even though the matter has non-linear inhomogeneities and complex physics such as  magneto-hydrodynamics is required to describe the evolution of at least the baryonic component of that matter? Even though certain neutrino species may currently be transitioning from the relativistic (radiation) regime to the non-relativistic weakly interacting (dust) regime?
    \item If so, is the average of the metric obtained from the full Einstein tensor (which is proportional to  the full stress-energy tensor) equal to the metric obtained from the Einstein tensor that is proportional to the average stress energy tensor?
    \item If the answer is ``yes, though not precisely'', how good is the approximation?
\end{itemize}
These questions themselves hide subtleties like -- what is meant by spatial average? Over which spatial hypersurfaces? The Einstein equations are local, and respect causality -- the spatial average of a quantity over some volume cannot causally determine the change of another quantity at any point until such time as the volume over which the average is taken is inside the past light cone of that point. In other words averaging requires abandoning local differential equations.

What can we imagine are the possible consequences of averaging on the evolution of the spatially averaged metric $\bar{g}_{\alpha\beta}$:
\begin{itemize}
    \item The average metric over the finite Hubble volume may not be precisely isotropic -- local anisotropies may average away slowly, or not at all. 
    \item The Friedmann equation obeyed by the average metric may not be represented to sufficient accuracy  by a sum of terms that are proportional to $a^{0}$, $a^{-2}$, $a^{-3}$, or $a^{-4}$. 
    In other words, \eqref{eqn:IntroduceOmegai} may not be a good representation of the evolution of the effective stress-energy, curvature, and cosmological-constant contributions to the right-hand-side of the Friedmann equation.
    \item $\Omega_i$ may not be sourced exclusively by the homogeneous sources to which they are ascribed -- for example there may be contributions to $\Omega_{-2}$ that are sourced by matter  non-linearities. 
\end{itemize}

There is another set of concerns relating to how we extract the values of $\Omega_i$ in \eqref{eqn:IntroduceOmegai}.   
In many cases this involves modeling the dynamics of the perturbations in the  metric and in the stress-energy  in the background metric.  
This is true for example in both CMB  and large-scale structure inferences of cosmological parameters.
Errors in the assumed form of \eqref{eqn:IntroduceOmegai} will propagate into the dynamical equations underlying those perturbations.

Thus to use \eqref{eqn:IntroduceOmegai} in the context of precision cosmology requires more than having a prior that favors a model for the stress energy that includes only radiation, baryonic matter, and non-interacting dark matter. 
It requires demonstrating that none of the concerns raised by the above questions lead to measurable deviations from the FLRW background model sourced by the average fluids.

As remarked above, Adamek {\textit et al.} \cite{Adamek:2018rru} provided numerical evidence supporting the claim (e.g., \cite{Kaiser:2015iia}) that, despite the presence of  non-linear structure, well-designed averages of observations could recover estimates of parameters in a background FLRW model
that are reasonably unbiased.
Specifically they found that a redshift-binned average of $1/d_L^2$ for Type 1a supernova luminosity distances  yields $\Omega_K$ consistent with the fiducial $\Omega_K=0$ universe, with a $1
\sigma$ uncertainty of  $\simeq0.001$.
However they analyzed only one simulation from which the precise value of neither the bias nor the theoretical variance in $\Omega_K$ can be inferred.
That study could resolve halos down to  $5\times10^{11}M_\odot$ -- how finer resolution (and greater non-linearity) or baryonic physics might affect the bias or variance remains uncertain.
Also uncertain is how inhomogeneities affect other cosmological observables from which FLRW background parameters are estimated.  
This has been addressed in some cases analytically in perturbation theory \cite{Bonvin:2015kea}, but not in the context of non-linear simulations.
Heinesen and Macphereson \cite{Heinesen:2020bej,Heinesen:2020pms,Macpherson:2021gbh,Heinesen:2021qnl} 
have worked to build a framework for further explorations of these matter.

One way to understand these issues is that the meaning of $\Omega_K$ may itself have an irreducible  ambiguity at the level of $\sim10^{-3}$, and so its value may well vary at this level among even well-designed observables that are used to infer ``it.''
As the precision with which experiments seek to infer $\Omega_K$, one should be cautious that this ambiguity not be misinterpreted as disagreement. 
On the other hand, this also suggests that properly designed estimates that result in values  larger than this ambiguity cannot be dismissed.
Truthfully, this caution extends to other potential parameters of FLRW cosmology that we quietly ignore, e.g. $\Omega_{-1}$ in the language of \eqref{eqn:Omegasubintegers}.  $\Omega_K$ stands out principally because it is an inherent  parameter of FLRW, i.e. it belongs to the geometric (left-hand) side of \eqref{eqn:EinsteinFieldEqns}, not (or not only, if we mean $\Omega_{-2}$) to the stress-energy (right-hand) side.

Finally, we must note that \cite{Adamek:2018rru} studied the effect of inhomogeneities only in a background $\Omega_K=0$ FLRW model. It is unclear how the results generalize to  $\Omega_K \neq 0$, or indeed to the wider set of allowed homogeneous but anisotropic models, or the even wider set of manifolds without, or with boundaries.  
A real program would seek to measure the full metric everywhere and then, if appropriate, fit it to a homogeneous metric plus perturbations.  While this is perhaps  an impossible program, we should be aware of the true physical problem and its challenges.

At this point we might be tempted to ignore that to describe our Universe we need a theory of the initial fluctuations.
Within vanilla-$\Lambda$CDM, this role is served by any one of a large set of possible inflation models, each with all of its attendant parameters. 
However,
proceeding boldly to ignore the questions of  inhomogeneities -- whether as initial conditions or generated in the very early universe ---
one could, instead of fitting for $\Omega_K$ (or better for fitting for a more general homogeneous background metric), opt to include a specification of the boundary conditions\footnote{Even manifolds without boundaries have boundary conditions -- e.g. the periodic boundary conditions that define a three-torus.}
for the homogeneous
background geometry as part of the model.
One would
then argue that a well-defined class of manifolds (with their associated boundary conditions) identifies a model,
and that we can do this in such a way as to focus our attention on FLRW models, and in particular flat-FLRW models.

To execute this boundary-condition approach, one must first assign a prior of $0$ to all manifolds with boundaries,  manifolds that do not admit a single homogeneous geometry, and manifolds that admit only an anisotropic geometry. 
One does this, despite the lack of any fundamental principle that excludes them, and in fact in the full knowledge that today we appear to live in a manifold with boundaries (black-hole singularities), or at least with regions hidden behind horizons.
One would be left with just the ``space'' of  manifolds that admit a homogeneous and isotropic geometry, i.e. the FLRW models.  
One would then, despite lacking a measure on this space, separate it into three classes -- those that admit homogeneous positive, zero, or negative curvature.
We could then call each class a model, and assign it a prior.
Homogeneous geometries in each class would have their own set of parameters that describe the metric\footnote{
    There will be others that characterize the different boundary conditions within each class -- e.g. the relative side lengths of the fundamental domain of a three-torus.
    As well, one will have some indexing of the different members of each model class, i.e. of the boundary conditions that all allow the same homogeneous geometry.}
The Euclidean metric has one fewer than the other two -- i.e. $\Omega_K\equiv0$.
One could then test 
whether the background curvature is flat, negatively or positively curved by performing model selection among the three models.
Usually this is done by fixing the boundary conditions within each class to select the covering space of that geometry.

This approach seems very hard to justify, especially given the absence of any theory that dynamically selects the homogeneous isotropic boundary-free manifolds from among all possible spaces,
or that chooses one of the Euclidean 3-spaces from among these manifolds.
Moreover, it is only through the fluctuations, for which one has adopted no theory, that one has observables through which to perform the model selection.

In the next section we explain why including inflation as the theory of the fluctuations prevents us from pursuing this program by generating through a random process non-primordial fluctuations that supplement (or replace) the  initial fluctuations.

\section{Inflationary Model Priors}

How does one justify assuming an FLRW background metric and then setting  $\Omega_K=0$, i.e. removing $\Omega_K$ from the set of free parameters of the theory, when no model yields $\Omega_K$ (or indeed the rest of $^3R_{ijkl}$)
exactly zero?
The standard answer is that inflation predicts that the geometry is currently (perturbed) FLRW, with $\vert\Omega_K\vert$ smaller than the uncertainty in any measurement we are likely to perform, justifying setting a $\delta(\Omega_K)$ prior\cite{Jaffeprivate}.

Certainly, there are inflationary models -- by which we mean Lagrangians of field theories coupled to GR\footnote{
    There must also be coupling to the Standard Model in order to effect ``reheating''--the transfer of a sufficient fraction of the potential energy density driving inflation into ordinary particles as inflation ends.}, as opposed to our previous use of model to denote a choice of background manifold --
that, for certain ranges of the Lagrangian parameters, take many cosmological initial conditions and map them into late-time universes in which certain spatial averages of the spatial Riemann tensor are small.
For those models, parameter choices and initial conditions, the flat-FLRW metric, and thus $\Omega_K=0$,  is an attractor of the classical theory  --- given any patch of the early universe in which the initial conditions allow inflation to begin and in which there is a short-wavelength cutoff to the initial perturbation spectrum, then, ignoring quantum fluctuations, the Riemann tensor in that patch will be driven toward that of the flat-FLRW metric, and $\Omega_K$ will be driven toward zero. 

This effective classical convergence to $\Omega_K$ indistinguishable from $0$ is not universal in inflationary models \cite{Efstathiou:2020wem}.
Certainly, if the inflation lasts forever, then one is driven classically toward a universe in which the only energy density is that of the inflation plus (any other source of) vacuum energy, and in which the spatial metric is precisely Euclidean.
On the other hand, if one is not in the right part of model parameter space or one has the wrong initial conditions, then inflation may not begin, or the Riemann tensor may not reach that of flat-FRW.
In ``just-so'' inflationary models, this leads to  $\Omega_K$ arriving near $0$, but to a value that may be observationally distinguishable from $0$. 
Meanwhile, there are also inflationary models that predict $\Omega_K\neq0$ for appropriate parameters \cite{Linde:1995xm,Hergt:2022fxk} -- such models were particularly popular when it was thought that $\Omega_K\neq0$ and are once-again becoming of interest.

Ironically, one should now realize that if inflation carries the scale of topology far beyond the Hubble scale, then it removes the possibility for an observer to use topology to determine which unique homogeneous geometry our spatial 3-manifold admits. 
Thus if we wanted to fix $\Omega_K=0$,
we would have to rely on a local-geometry meaning of $\Omega_K$.
Here the fact that inflation is used not just as an explanation for the near-homogeneity, near-isotropy and near-flatness of the universe, but also as an explanation for its inhomogeneities is crucial, as we now discuss.

The great success of the inflationary scenario is widely regarded as the prediction of a spectrum of scalar quantum fluctuations that (at least by the time they are observed $14$ Gyr later): are  predominantly or entirely adiabatic; are nearly, but not precisely, scale free; extend at least to scales that until recently were super-horizon; and are nearly Gaussian at early times.
These quantum fluctuations are produced at sub-horizon scales in an ongoing process during inflation even as the initial fluctuations in the stress-energy and geometry are being stretched to ever-longer and eventually super-horizon physical scales.
These quantum fluctuations carry components of the Riemann tensor away from zero even as the accelerated expansion of the Universe is redshifting the initial fluctuations and their contribution to the Riemann tensor.

To summarize,
the background initial conditions pre-inflation could be $\Omega_K=0$ (in the topological sense), but long-enough inflation 
removes the ability of the observer to use the topology to anchor a unique choice of homogeneous geometry, and thereby set $\Omega_K$ exactly to $0$.
Inflation simultaneously generates fluctuations in the metric which 
create  an irreducible ambiguity for any observer about the precise value of $\Omega_K$ in their Hubble patch -- different ``nearby'' choices of homogeneous background geometry will yield ``nearby'' descriptions of the fluctuations.
All of them will be consistent with the theory--
there is no 
unique way to perform the background-plus-perturbations split. 
Hence we cannot assume $\Omega_K=0$, with a theory that can only describe $|\Omega_K|\simeq 0$;
but how ``nearby'' does inflation predict these descriptions to be, i.e. does inflation have a generic prediction of the amplitude and spectrum of post-inflationary fluctuations? 

When inflation was first proposed, it was expected that the amplitude of the quantum-induced scalar fluctuations would naturally be ${\cal O}(1)$ -- and that in most inflationary models a fine-tuning of model parameters would be necessary to suppress the fluctuations, so as not to spoil the inflationary prediction of homogeneity and flatness.
As the search for CMB fluctuations continued, it  resulted in increasingly stringent upper limits on their amplitude,
so  the  required fine tuning became increasingly stringent.
The scalar fluctuations were finally detected by the Cosmic Background Explorer \cite{COBE:1992syq}, but at a level that 
requires either stringent fine tuning or clever models in which the necessary small parameters emerge ``naturally.''
The original {\it a priori} space of  parameter of inflationary models was slowly shrunk by gradually improving observational limits, with the result that what we confidently believed about inflationary models to begin with was not matched by our posterior inferences.\footnote{
    Of course every improved measurement of a fundamental parameter shrinks the allowed parameter space, and one would not want to view that as generically problematic for a theory. 
    This makes the precise characterization of fine-tuning difficult; however, if dimensionless parameters of models are forced to have very small values, or very particular ratios without an explanation through symmetry, that is generally viewed as concerning. }
Moreover, this evolution to more restricted parts of parameter space -- in which FLRW would be a better approximation -- was driven by observations, not new understanding of theoretical principles.  

There have been concerted efforts to demonstrate that there is a large set of initial conditions in which inflation begins and persists for long enough to homogenize, isotropize, and flatten the Universe.
Arguments have been made that the appropriate measure on the space of initial conditions favors those that support inflation, and that the more inflation they support they more heavily they should be weighted.
There has also been a  history of cosmologists remaining or becoming unconvinced by the explanatory power of this mapping. 
Priors  on  possible initial conditions are difficult to assess, 
and the on-again-off-again interest in the ``measure problem'' for inflationary models is probably a reflection of this difficulty.
(See, e.g. \cite{Vanchurin:1999iv,Garriga:2005av,Bousso:2006ev,Guth:2007ng,Linde:2008xf,Linde:2010xz,Guth:2011ie}.)

The tensor fluctuations from inflation have not yet been detected as ``B-modes'' in the cosmic microwave background.
That non-detection has itself begun driving not just a shrinking of the parameter spaces of inflationary models, but a  shrinking of the allowed space of inflationary models. 
Priors on the set of possible inflationary models are even more difficult to assess than the priors on model parameters.
However, over the last decades  many of the single-field inflationary models that were regarded as heavily favored \textit{a priori} are now effectively ruled out.
Yet, the claim that a very narrow prior on $\Omega_K$ is justified rests on confidence in our ability to pre-judge inflationary models.

Should we have such confidence in an unspecified and shrinking set of inflationary models that we insist that an observable cosmological parameter $\Omega_K$ should not be estimated because in some of those models, with certain parameter choices (and some range of initial conditions) that have been dictated by observations, its expected value cannot currently be distinguished from zero? 

More concisely, within the inflationary framework, one is not meant to include a specification of the boundary conditions for the homogeneous and isotropic background as part of the model, the FLRW background geometry is meant to emerge dynamically. 
Meanwhile,  the theory for the inhomogeneous part of the metric, make it impossible to expect that the average curvature, and hence $\Omega_K$, is exactly zero.
Instead, the observed properties of fluctuations is what drives limits on inflationary models and their parameters.
Perhaps one could try to argue that it is possible to test $\Omega_K \simeq 0$ FLRW against the data. However this choice of priors is not motivated by our theoretical understanding, i.e.~there are no theoretical reasons to select initial conditions patches and inflationary model parameter values to fulfill the $\Omega_K \simeq 0$ FLRW condition and then call this a ``cosmological model''. 
Only data can tell us if our inflationary-$\Lambda$CDM description works with FLRW geometry and $\Omega_K$ consistent with zero or not. Hence, even within an FLRW assumption, the value of $\Omega_K$ needs to be properly inferred from observational data.

\section{Some practical takeaways regarding spatial curvature}

The universe is not precisely homogeneous and isotropic, it has fluctuations.
$\Omega_K$, and for that matter the other cosmological parameters of FLRW, are phenomenological parameters, not fundamental parameters.  
They describe a background universe that doesn't actually exist, but which is meant to approximate the Universe (or the piece of it in which we live) well enough that we can treat deviations from FLRW as perturbations with statistics we can relate simply to our model of the origin and evolution of those perturbations.
Certain inflationary models, with certain values of their parameters, map certain parts of the space of initial conditions to regions of spacetime in which spatial curvature is small but not exactly zero, but our track record in reliably placing strong priors on this extremely complicated space of inflationary models, parameters and initial conditions does not justify placing strongly informative priors on cosmological parameters and then claiming that this set of priors defines a new ``cosmological model'' named flat-$\Lambda$CDM where the background is homogeneous, isotropic, and Euclidean 
($\Omega_K = 0$).  

One should not fix an FLRW parameter to a specific value without a theoretical framework to enforce that value. It is not appropriate to regard such fixing as ``a restrictive assumption that must be tested empirically'' \cite{Planck:2015fie}, if that means demanding strong Bayesian evidence to negate the assumption and restore the parameter.
In particular, there is no compelling theoretical reason to set $\Omega_K=0$. 
It must be included in the set of FLRW that are to be fit.

If one finds that $\Omega_K\neq0$, then it will be important to explore whether other degrees of freedom of the spatial curvature ($^3R_{ijkl}$) are non-zero.
This is necessary to understand the cause of the non-zero value of $\Omega_K$, in particular  whether the deviation is consistent with the universe being nearly FLRW.

Notice (e.g. \cite{Wright:2006kx,Clarkson:2007bc,Planck:2018vyg,DiValentino:2019qzk,Handley:2019tkm}) that fitting for $\Omega_K$ rather than fixing it has consequences for best-fit values and errors of other cosmological parameters, and for assessments of the consistency of cosmological data sets (e.g.~ \cite{Gonzalez:2021ojp}).

Theorists should also consider predicting and observers then measuring quantities that have clear meanings not just in the background metric but in the actual metric (and that may reduce to simpler quantities in a homogeneous and isotropic limit).   For example, components of the ($3+1$-d) Riemann or Ricci tensors, or the Ricci scalar.

\section{Summary}

We examined the common practice among cosmologists of omitting $\Omega_K$  from the set of 7 standard FLRW parameters to be fit to data, by setting it precisely to $0$, 
and then arguing that to add it requires sufficient  evidence that it is not $0$.
This practice is given the name ``flat $\Lambda$CDM,'' suggesting that it is a well-defined separate and fundamental theory of the Universe from curved $\Lambda$CDM.
However, the 7 parameters of inflationary $\Lambda$CDM cosmological theory are phenomenological parameters of an approximate effective theory of the Universe on large, but not necessarily superhorizon, scales.  
The map to them from the much larger number of parameters of an ultimate complete underlying microscopic theory -- General Relativity, plus the Standard Model (SM) of particle physics, plus any Beyond the SM fundamental physics ultimately needed to account for dark matter, dark energy, inflation, baryogenesis or leptogenesis, the quantum nature of gravity,  and whichever current observational anomalies ultimately prove to be physical -- is unknown, but undoubtedly highly nonlinear, and probably chaotic in many parts of the parameter space of the fundamental theory. 
It may well also involve a specification of a wide variety of data on some Cauchy surface; or perhaps it has no deterministic specification.

Since the injection of this information, $\Omega_K\equiv0$ {\em a priori} may alter the inferred values of the 6 other parameters and will reduce the uncertainty on them, we need to understand the underlying justification for this delta-function prior on a phenomenological parameter.

We began by exploring the meaning of the term flat and its connection to any arguments underlying the prior.
We noted that a spatial hypersurface  on which we find ourselves (``space'') could be a manifold that admits a unique homogeneous geometry, and that could be the Euclidean geometry.
It is extremely challenging to justify any specific measure on the space of manifolds, and to use this to argue convincingly for a strong preference for a topologically flat manifold. 
A compelling theory of quantum cosmology might help in this regard, but it is currently lacking.
The topology of space thus seems destined to be a question that can be answered only by observation, if it is answerable at all.
Cosmic topology can in principle be determined from appropriate measurements if the spatial hypersurface is  topologically non-trivial and the relevant length scales are not too superhorizon;  efforts to search for such topology has so far yielded only lower limits on certain associated observational length scales. 

We therefore have no theoretical justification to choose, out of all possible manifolds, those manifolds that admit a Euclidean metric and declare them to be a separate privileged (high-prior) class.
 
A second, more relevant definition of flatness is that appropriate averages of the Riemann tensor of the 3-geometry of space are zero. 
We pointed to some of the many questions or challenges surrounding this definition. 
Perhaps salient among them is that we have had very little observational access to any sizable sub-volume of space at a fixed time -- our information about the distant reaches of the Universe comes to us at the speed of light, and so we observe on our past light cone.
\footnote{We will eventually get some information from observables like the Sunyaev-Zeldovich (SZ) effect where light is scattered into our past light cone, and also from messengers, like neutrinos and cosmic rays, that are subluminal.}
We could in principle make observations of the geometry and the stress energy on that past light cone and evolve it forward to the current time, but even in principle that would only determine the geometry here, since every other location in space has points on its past light cone that are outside our past light cone. 
We are thus limited, perhaps in principle, to checking consistency of observations with a theory that predicts the statistical properties of the spatial Riemann tensor on our past light cone.

We do have a theory that makes predictions about the behavior of the spatial Riemann tensor -- inflation.
At least in regimes where deviations from a flat FLRW spacetime are linear, the primary effect of inflation on a classical level is to change the relation between comoving and physical scales, so that short-wavelength fluctuations become long-wavelength.  
Under the assumption that the initial conditions have no fluctuations below some cutoff scale, this will cause the Riemann tensor measured over a fixed physical volume to eventually approach that of a Euclidean cosmological spacetime.
This stretching out of the fluctuations is balanced by quantum mechanical generation of new fluctuations on small scales (which are then also stretched) which can be arranged to be of small amplitude.
At the end of inflation, the nearly homogeneous energy density driving inflation could be transformed efficiently into ordinary matter and radiation, and the Riemann tensor could be close to that of $k=0$ FLRW -- spatially nearly homogeneous and nearly isotropic, with nearly zero 3-Riemann tensor. 
Whether or not inflation begins, and how closely this classical evolution would cause the patch we are in to approach the Euclidean LFRW geometry, depends on the initial pre-inflationary geometric inhomogeneities, on the parameters of the inflationary theory, on the initial conditions of the inflaton field, and on other specifics of the inflationary model.

For a very long time after that post-inflationary thermalization, or perhaps forever -- depending on the underlying topology, and the magnitude of vacuum energy (or the value of $\Lambda$), and amplitude of the quantum fluctuations -- the Universe would be well described approximately by an FLRW metric with $k=0$.  Its dynamics would therefore be close, by some measure, to those of an $\Omega_K=0$ FLRW universe.

Predictions of a small current value of $\Omega_K$ -- i.e. below the observational uncertainties of our current experiments -- are  thus inflationary-model dependent, model-parameter dependent, and (to a widely debated extent) initial-condition dependent. 
We have to be in an inflationary model that has a flat-FLRW attractor;  we have to be in a region of initial condition ``space'' that is in the basin of that attractor;  the quantum fluctuations cannot drive us too far away from the FLRW Riemann tensor on large scales;  the non-linear  fluctuations on small scales must average in some precise way to recover a Hubble-scale flat-FLRW Riemann tensor with some appropriate observational scheme. 

Our ability to predict \textit{a priori} which is the correct inflationary Lagrangian has already been shown to be poor. We have even  less theoretical justification for choosing the values of the associated Lagrangian parameters, the initial inflaton field value, the initial metric and inflaton inhomogeneities, and whatever else  determines the duration of inflation and the amplitude of inflationary fluctuations.
These are determined from observation, not theoretical principles.
Since $\Omega_K$ depends on these, we cannot justify an $\Omega_K=0$ prior, or even a prior that enforces that $\vert\Omega_K\vert$ be less than our observational error bars.

Setting $\Omega_K=0$ does not define a ``cosmological model'', it is just an arbitrary insertion of strong information into FLRW parameter determination.
Within FLRW, $\Omega_K$ must be estimated from observations.

\section{Acknowledgements}
GDS is partly supported by a Department of Energy grant DESC0009946 to the particle astrophysics theory group at CWRU. 
SA is party supported by the project ``Combining Cosmic Microwave Background and Large Scale Structure data: An Integrated Approach for Addressing Fundamental Questions in Cosmology,'' funded by the MIUR Progetti di Rilevante Interesse Nazionale (PRIN), Grant Number 2017YJYZAH, Bando 2017.
JBM is partly supported
by a Department of Energy grant DE-SC0017987 to the
particle astrophysics theory group at WUSTL.
J. T. G. is supported by the National Science Foundation, Grant No. PHY-2013718.
We thank Y. Akrami, C. Copi, A. Jaffe, and A. Kosowsky for enlightening discussions.

\bibliography{references.bib}
 
\end{document}